\documentclass[aps,prd,superscriptaddress,12pt,showpacs,notitlepage]{revtex4}
	\usepackage{graphicx}
	\usepackage{amsmath,amssymb,mathrsfs}
\usepackage{epsf}
\usepackage{epsfig}
\usepackage{subfigure}
\usepackage{amssymb}
\usepackage{amsmath}

\newcommand{\be}{\begin{equation}}
\newcommand{\ee}{\end{equation}}
\newcommand{\bea}{\begin{eqnarray}}
\newcommand{\eea}{\end{eqnarray}}

\newcommand{\bb}{\bibitem}
 
\newcommand{\eqn}{\begin{eqnarray}}
\newcommand{\eqnx}{\end{eqnarray}}

\begin{document}
\title{Composite BPS skyrmions from an exact isospin symmetry breaking}

\author{Pawel Klimas}
\affiliation{Universidade Federal de Santa Catarina, Trindade, 88040-900, Florian\'o—polis-SC, Brazil}
%\author{C. Naya}
%\affiliation{Departamento de F\'isica de Part\'iculas, Universidad de Santiago de Compostela and Instituto Galego de F\'isica de Altas Enerxias (IGFAE) E-15782 Santiago de Compostela, Spain}
%\author{J. Sanchez-Guillen}
%\affiliation{Departamento de F\'isica de Part\'iculas, Universidad de Santiago de Compostela and Instituto Galego de F\'isica de Altas Enerxias (IGFAE) E-15782 Santiago de Compostela, Spain}
%\author{A. Wereszczynski}
%\affiliation{Institute of Physics,  Jagiellonian University,
%Reymonta 4, Krak\'{o}w, Poland}

%\pacs{11.30.Pb, 11.27.+d}

\begin{abstract}
We study the BPS Skyrme model with potentials breaking the isospin symmetry and analyse how properties of exact 
solitonic solutions depend on a form of the isospin breaking potential. In the case of the strong symmetry breaking a new topologic structure is observed which enables us to decompose a BPS skyrmion into a lower dimensional defect localised 
on a brane (kink). 
We investigate some thermodynamical properties of such solitons as well as the role of the symmetry breaking potential in the resulting mean-field equation of state. 
\end{abstract}

\maketitle 
%%%%%%%%%%%%%%%%%%%%%%%%%%%%%%%%%%%%%%%%%
\section{Introduction}
%%%%%%%%%%%%%%%%%%%%%%%%%%%%%%%%%%%%%%%%%
The Skyrme model \cite{skyrme} is a widely accepted candidate for the effective model of the low energy QCD. It is a purely mesonic field theory ($U\in SU(2)$) where baryons and atomic nuclei are not introduced as primary fields but emerge as nonperturbative excitations {\it i.e.}, topological solitons with an identification between topological index $Q \in \pi_3(\mathbb{S}^3)$ and the baryon charge $B$ \cite{thooft}. 

After the semiclassical quantization \cite{nappi} the model describes not only baryons (proton, neutron, $\Delta$) but also spectra of light nuclei \cite{nuclei}, \cite{halcrow}. Quite recently it has been also understood how to modify the original proposal in such a way that it can be used in nuclear physics. This requirs to bring the model close to a BPS theory which resultes in a significant reduction of binding energies of atomic nuclei. Furthermore, such a model should describe a fluid matter rather than a crystal as it is the case in the simplest Skyrme model. All this can be achieved by addition to the minimal model
\be
\mathcal{L}_{Sk}=\tilde{\mathcal{L}}_0+\mathcal{L}_2+\mathcal{L}_4,
\ee
\begin{equation}
\mathcal{L}_2=-\lambda_2 \mbox{Tr}\; (L_\mu L^\mu), \;\;\; \mathcal{L}_4=\lambda_4 \mbox{Tr} \; ([L_\mu , L_\nu]^2), \;\;\;  L_\mu \equiv U^\dagger \partial_\mu U
\end{equation}
a BPS part (the {\it BPS Skyrme model} \cite{BPS})
\begin{equation} \label{BPSmodel}
\mathcal{L}_{BPS} \equiv \mathcal{L}_6+\mathcal{L}_0 \equiv -(24\pi^2)^2 \lambda_6 \mathcal{B}_\mu \mathcal{B}^\mu- \lambda_0 \mathcal{U}.
\end{equation}
It consists of the sextic term $\mathcal{L}_6$ - the square of the baryon current 
\begin{equation} \label{top-curr}
\mathcal{B}^\mu = \frac{1}{24\pi^2} \epsilon^{\mu \nu \rho \sigma} \mbox{Tr} \; (L_\nu L_\rho L_\sigma), \quad B\equiv \int d^3 x \mathcal{B}^0\
\end{equation}
($B$ is baryon number) - and a further potential $\mathcal{U}$, which is assumed not to change the pion mass provided by the the potential $\tilde{\mathcal{L}}_0$ from the usual Skyrme part. 
The main observation is that in the exact BPS limit (the BPS Skyrme model) one finds a theory which has zero binding energies (at the classical level) and which is a field theoretical realization of a perfect fluid \cite{term} - a natural state of matter in nuclear physics. Non-zero physical binding energies can be obtained again by the semiclassical quantization procedure and by taking into account the Coulomb interaction as well as the isospin symmetry breaking \cite{bind} (see also \cite{marl}). Let us notice that another possibly which reduces classical binding energies of skyrmions to the physical values {\it i.e.}, the loosely bound Skyrme model \cite{loos}, can be in a sense included into the near BPS model. In this approach low binding energies are obtained by a suitable modification of the potential - of course, without spoiling the pionic masses. But this potential can be naturally included into the BPS part of the model. On the other hand, the sextic term is unavoidable to make a close relation to a perfect fluid. In any case a particular form of the potential have a strong impact on physical properties of skyrmions ({\it e.g.} binding energies). Therefore, it is important to further study the role of the potential in the Skyrme type theories and its impact on geometrical features of the solitons. 

Although the complete description of nuclear physics phenomena is expected to be given by the full near-BPS Skyrme model 
\begin{equation}  \label{full}
\mathcal{L}=\mathcal{L}_{Sk} + \mathcal{L}_{BPS}, 
\end{equation}  
and is probably a mutual effect of inclusion of the sextic term (BPS model) and properly chosen potential (loosely bound model)
some (but certainly not all \cite{time}) features are governed by the BPS part of the model. It is due to the physical relevance of the BPS limit for nuclear matter, which makes it reasonable to expect that the masses of skyrmions {\it i.e.}, atomic nuclei, should be mainly provided by the BPS part of the model. Hence, the BPS limit can be still interesting and relevant for some bulk observables. This is important since the BPS Skyrme model is a solvable solitonic theory. It provides an analytical insight into physical properties of higher nuclei, nuclear matter and neutron stars. For example, the solvability of the model allows for a {\it beyond} mean-field theory description of neutron stars \cite{stars} (for gravitating BPS skyrmions see also \cite{bjarke}).
   
The main aim of the present paper is to further investigate some classical aspects of the BPS Skyrme model and, in particular, to  analyze how properties of solutions are modified if one choses a potential which breaks the isospin symmetry. 

First of all we break the isospin symmetry in a rather strong manner. The reason for that is that we want to qualitatively understand how skyrmions reflects the form of the potential rather than to compute some isospin breaking effects in the nuclear physics. Such strong symmetry breaking potentials can deform the target space geometry and lead to new type topological defects. In particular, we want to check if it is possible to decompose BPS skyrmions into a lower dimensional defect localised on a brane (domain wall) - as discussed recently in \cite{nitta1}-\cite{jen}. Obviously, such a target space deforming symmetry breaking has no direct application to nuclear physics but it is an interesting mathematical phenomenon which can be analytically studied in the BPS Skyrme model. However, our framework allows to perform an analytical study of weak isospin symmetry breaking as well. Such a symmetry breaking possesses a direct application to nuclear physics if the semiclassical quantisation is taken into account. 

Secondly, we want to understand how the (weak and strong) isospin symmetry breaking change thermodynamical properties of  matter described by skyrmions. Specifically, we shall study some modifications of the equation of state caused by the isospin symmetry breaking.

%%%%%%%%%%%%%%%%%%%%%%%%%%%%%%%%%%%%%%%%%
\section{The exact isospin symmetry breaking}
%%%%%%%%%%%%%%%%%%%%%%%%%%%%%%%%%%%%%%%%%
The isospin symmetry breaking in the framework of the (BPS) Skyrme model is accomplished by an assumption that the potential gives different masses for the pionic, perturbative excitations. As usual the pions $\vec{\pi}$ are introduced as components of the mesonic matrix field
\be
U= \sigma \mathbb{I} + i\vec{\pi}\cdot  \vec{\tau}
\ee 
where we have the constraint
\be
\sigma^2 + \vec{\pi}^2=1.
\ee
Hence, if the potential depends only on $\sigma$, that is in the same way on each component of the pionic degrees of freedom, the isospin symmetry is present. In the paper we assume a different parametrisation of the matrix field, namely
\begin{equation}
U(x)=e^{i\xi \,\vec{n} \cdot \vec{\tau}}  
\end{equation}
where $\xi=\xi(x)$ is a scalar (profile of skyrmion) and $\vec{n}=\vec{n}(x)$ is a unit  three component vector field while $\vec{\tau}$ are Pauli matrices and $x\equiv(x^0,\ldots,x^3)$. Obviously, $\vec{\pi} = \sin (\xi)\, \vec{n} $ and $\sigma =\cos (\xi)$. 
For convenience reason we also use the standard stereographic projection which expresses $\vec{n}$ by a complex scalar field $u$
\begin{equation}
\vec{n}=\frac{1}{1+|u|^2} \left( u+\bar{u}, -i(u-\bar{u}), 1-|u|^2\right).
\end{equation}
Using our parametrisation we can rewrite the BPS Skyrme Lagranian as
\begin{equation}
\mathcal{L}=\frac{\lambda^2\sin^4\xi}{(1+|u|^2)^4} \left( \epsilon^{\mu \nu \rho \sigma} \xi_\nu u_\rho \bar{u}_\sigma \right)^2 -\mu^2 \mathcal{U}.\label{Lagr}
\end{equation}
Here $\lambda_6 \equiv\lambda^2/(24)^2$ and $\lambda_0\equiv\mu^2$ with $\lambda>0$ and $\mu>0$. The espression $\epsilon^{\mu \nu \rho \sigma}$ in (\ref{top-curr}) and (\ref{Lagr}) stands for the Levi-Civita tensor $ \epsilon^{\mu \nu \rho \sigma}=-\frac{1}{\sqrt{-g}}\epsilon'^{\mu \nu \rho \sigma}$ where $\epsilon'^{\mu \nu \rho \sigma}\equiv\epsilon'_{\mu \nu \rho \sigma}$ is the  antisymmetric  symbol with $\epsilon'_{0123}=+1$.  We have also adopted a short-hand notation $\xi_{\nu}\equiv\frac{\partial\xi}{\partial x^{\nu}}$ and $u_{\rho}\equiv\frac{\partial u}{\partial x^{\rho}}$. The BPS model revels the isospin symmetry breaking if the potential $\mathcal{U}$ depends not only on the profile function $\xi$ but it possesses also a part which is a function of the complex field
\be
\mathcal{U}=\mathcal{U}(\xi, u,\bar{u}).
\ee
The situation is especially simple if the potential factorises into two terms
\be
\mathcal{U}=V(\xi) W(u\bar{u}).\label{factor}
\ee
Below we shall explore some situations where the potential factorises as in (\ref{factor}).
%%%%%%%%%%%%%%%%%%%%%%%%%%%%%%%%%%%%%%%%%
\subsection{The BPS equations}
%%%%%%%%%%%%%%%%%%%%%%%%%%%%%%%%%%%%%%%%%
It is known that the BPS Skyrme model admits a reduction of the static field equations to 
a BPS equation \cite{BPS}, \cite{Speight1}, \cite{stepien}. Namely, the energy takes the form
\bea
E&=&\int d^3 x \left( \pi^4\lambda^2 \mathcal{B}_0^2  +\mu^2 \mathcal{U}\right) \nonumber\\
&=& \int d^3 x   \left( \pi^2\lambda \mathcal{B}_0  \pm \mu \sqrt{\mathcal{U}} \right)^2 \mp 2\pi^2\lambda \mu \int d^3 x \mathcal{B}_0 \sqrt{\mathcal{U}} \nonumber\\
&\geq&  2\pi^2 \lambda \mu \left| \int d^3 x \mathcal{B}_0 \sqrt{\mathcal{U}} \right| = 2\pi^2 \lambda \mu |B| \left< \sqrt{\mathcal{U}} \right>_{\mathbb{S}^3},
\eea
where the average of the potential on the target space has been introduced
\be
\left< \sqrt{\mathcal{U}} \right>_{\mathbb{S}^3} \equiv \frac{1}{2\pi^2} \int \mbox{vol} _{\mathbb{S}^3} \sqrt{\mathcal{U}}.
\ee
The bound is saturated by solutions of the Bogomolnyi equation
\be
 \pi^2\lambda \mathcal{B}_0  \pm \mu \sqrt{\mathcal{U}} =0,\label{Bogomolny}
\ee
which are also solutions of the full static field equations. Note that $\mathcal{B}_0=\mathcal{B}^0$ in adopted here convention $(+,-,-,-)$ for signature of the metric tensor. For analysed here the isospin symmetry breaking potential the Bogomolnyi equation can be further decomposed into two decoupled first order equations. The  Bogomolnyi equation reads
\begin{equation}
i \lambda\frac{ \sin^2 \xi}{(1+|u|^2)^2} \frac{1}{\sqrt{g_3}} \epsilon_{abc}'  \xi_a  u_b  \bar{u}_c= \pm \mu \sqrt{V} \sqrt{W},\label{bps0}
\end{equation}
where $a,b={1,2,3}$ and $g_3$ stands for determinant of a metric tensor in a 3-dim Euclidean space. 
Equation (\ref{bps0}) is decomposed as follows
\begin{equation}
i\lambda \frac{1}{\sqrt{g_2}} \epsilon_{ij}' \frac{ u_i \bar{u}_j}{(1+|u|^2)^2}=\eta_1  \alpha \sqrt{W},\label{bps1}
\end{equation}
\be
\sin^2 \xi \frac{1}{\sqrt{g_1}} \xi_k=\eta_2 \frac{\mu}{\alpha} \sqrt{V},\label{bps2}
\ee
where $x^k$ is a chosen single coordinate ($k$ - fixed) different form $x^i,\,x^j$ and signs $\eta_1,\eta_2 =\pm 1$ have been chosen in a way that $\eta_1\eta_2=\pm 1$ is by definition equal to the sign on rhs of (\ref{bps0}). It follows from separation of coordinates that $\sqrt{g_3}=\sqrt{g_1}\sqrt{g_2}$ where $g_2$ is determinant of a metric tensor on a surface parametrized by coordinates $x^i$, $i\neq k$ and $\sqrt{g_1}$ is a Lam\'e coefficient associated with a third one coordinate $x^k$. An arbitrary constant $\alpha$ has been introduced in order to find solutions which cover full target space and has nontrivial topological charge.  
Now, if we assume the following ansatz for the scalar real and complex functions
\be
\xi = \xi (x^k), \;\;\; u=u(x^i, x^j),
\ee
then the decomposed Bogomolnyi equations not only imply the full Bogomolnyi equation but also the original second order equations of motion and therefore lead to solutions of the isospin symmetry broken BPS Skyrme model. 

Such decomposition of the Bogomolnyi equation leads to decomposition of the energy bound and the baryon topological charge. Indeed, 
\bea
E &\geq& \mp2\pi^2\lambda\mu\int d^3 x \mathcal{B}_0  \sqrt{\mathcal{U}} \nonumber\\ &=&\pm 2 \lambda \mu  \int d^3 x\, \frac{i}{\sqrt{g_3}} \epsilon'_{ij} \frac{ u_i \bar{u}_j}{(1+|u|^2)^2} \sin^2 \xi \xi_k  \sqrt{W} \sqrt{V} \nonumber\\
&=& \pm 2 \lambda \mu \left( \int d^2 x\, \frac{i}{\sqrt{g_2}}\epsilon'_{ij} \frac{ u_i \bar{u}_j}{(1+|u|^2)^2} \sqrt{W}  \right) \cdot \left( \int dx^k \sin^2 \xi \xi_k  \sqrt{V} \right) \label{bound1} \nonumber \\
&=& \pm2 \lambda \mu \;  Q_{\mathbb{S}^2}  \left< \sqrt{W} \right>_{\mathbb{S}^2} \left< \sqrt{V} \right>\label{bound}
\eea
where
\begin{eqnarray}
Q_{\mathbb{S}^2} = \frac{i}{2\pi} \int d^2 x \epsilon'_{ij} \frac{\nabla_i u \nabla_j \bar{u}}{(1+|u|^2)^2} \label{Q2}
\end{eqnarray}
is the topological charge of the baby skyrmion $Q_{\mathbb{S}^2} \in \pi_2(\mathbb{S}^2)$. The symbol $\nabla_i$ stands for gradient components on the unit sphere. The mean values are given by integrals
\be
\left< \sqrt{W}\right >_{\mathbb{S}^2} \equiv \frac{1}{4\pi} \int_{\mathbb{S}^2} \mbox{vol}_{\mathbb{S}^2}  \sqrt{W},  
\ee
\be
\left< \sqrt{V} \right> \equiv \int dx^k\xi_k \sin^2 \xi   \sqrt{V}. 
\ee
Obviously, the usual non-symmetry breaking case {\it i.e.} $W=1$ is also included into this decomposition. Then, assuming the spherical coordinates $(r,\theta,\phi)$ we have $\xi=\xi (r)$ and $u=u(\theta, \phi)=v(\theta) e^{in\phi}$. The integer number $n$ has interpretation of baryon number $n\equiv B$. The pertinent solution of $W$ part reads
\be
u(\theta,\phi)=\tan\left( \frac{\theta}{2}\right) e^{in\phi}\label{u0}
\ee
and requires $\alpha = \eta_1n\lambda /2$. Finally, the profile is determined by the equation
\be
\frac{1}{r^2} \sin^2 \xi\, \frac{d\xi}{dr} =\eta_1\eta_2 \frac{2\mu}{n \lambda} \sqrt{V}.
\ee
It became clear from the above analysis that the energy bound of BPS skyrmions and the BPS equations, 
can be naturally decomposed into a collection (superposition) of lower dimensional defects which 
carry pertinent topological charges. This is quite similar to cases discussed in \cite{nitta1}-\cite{jen}, although the
action is different. 

%%%%%%%%%%%%%%%%%%%%%%%%%%%%%%%%%%%%%%%%%
\section{Examples} 
%%%%%%%%%%%%%%%%%%%%%%%%%%%%%%%%%%%%%%%%%

%%%%%%%%%%%%%%%%%%%%%%%%%%%%%%%%%%%%%%%%%
\subsection{Baby skyrmions on a spherical brane}\label{sec2}
%%%%%%%%%%%%%%%%%%%%%%%%%%%%%%%%%%%%%%%%%
In this an next two sections we shall deal with decomposition  which lead to some composite structures containing as ingreedients baby-skyrmions {\it i.e.} solutions of the Skyrme model in lower dimensions.  Such low-dimensional solutions of the Skyrme model have already been investigated in literature in the case where physical space is either Euclidean two dimensional space or a Minkowski (2+1) spacetime, see for instance \cite{wojtek}.

\begin{figure}[h!]
\begin{center}
\includegraphics[width=0.45\textwidth]{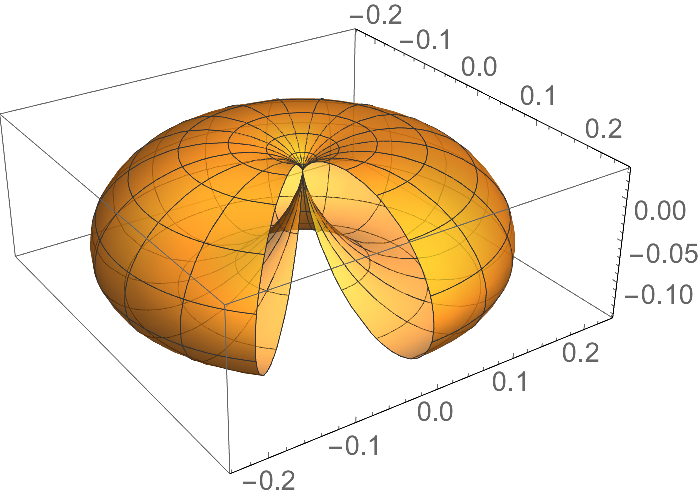}\caption{The energy density contribution $W(\theta)\equiv W(u(\theta,\phi)\bar u(\theta,\phi))$. The distance from the center represents the value of $W(\theta)$.}\label{don}
\end{center}
\end{figure}

We shall study a potential which breaks the isospin symmetry, {\it i.e.} it is of the form $\mathcal{U}=V(\xi)W(u\bar u)$, where
\begin{equation}
V=1-\cos \xi, \;\;\;\; W=  \frac{|u|^2}{1+|u|^2} = \frac{1}{2}(1-n^3).
\end{equation}
For axially symmetric ansatz $\xi =\xi(r)$, $u=u(\theta, \phi)=v(\theta)e^{in\phi}$ the BPS equation can be cast in the form of two ordinary  differential equations  
\bea
&& 2\lambda n\frac{1}{\sin\theta}\frac{v\,v_{\theta}}{(1+v^2)^{2}}=\eta_1\alpha \frac{v}{\sqrt{1+v^2}},\label{bps1b}\\
&&\frac{1}{r^2}\sin^2\xi\,\xi_r=\frac{\eta_2}{\alpha}\mu\sqrt{1-\cos\xi},\label{bps2b}
\eea
where $\xi_r\equiv\frac{d\xi}{dr}$ and $v_{\theta}\equiv\frac{dv}{d\theta}$.
The system of equations (\ref{bps1b}) and (\ref{bps2b}) has a constant solution $v=0$ and $\xi=0$. In order to find a non-trivial solution of (\ref{bps1b}) we choose $\alpha=\eta_1\lambda\, n$, then
\be
v(\theta)=\frac{\tan\frac{\theta}{2}\sin\frac{\theta}{2}}{\sqrt{1+\sin^2\frac{\theta}{2}}}.\label{v1}
\ee
This solution vanishes at $\theta=0$ and tends to infinity as $\theta\rightarrow\pi$. The solution of (\ref{bps2b}) became a compacton
\begin{equation}
\xi(r) = \left\{ 
\begin{array}{ll}
2\, \mbox{arccos} \;\left( \frac{r}{R_0}\right) & r \leq R_0 \\
0 & r > R_0
\end{array}
\right.
\end{equation}
whose radius $R_0$ takes the value 
\begin{eqnarray}
R_0 = \sqrt[3]{\frac{4\sqrt{2} \lambda |n|}{\mu} },
\end{eqnarray}
where in order to get $R_0>0$ one has to choose $-\eta_1\eta_2\,{\rm sign}(n)=1$. It means that solution with $n>0$ $(n<0)$ satisfies (\ref{bps0}) with $"-"$ $("+")$.   The solution saturates the energy bound 
\be
E=2\eta_1\eta_2Q_{\mathbb{S}^2}\left<W\right>_{\mathbb{S}^2}\left<V\right>=\frac{128\pi}{45}\sqrt{2}\lambda\mu |n|,\label{energia1}
\ee
where
\bea
Q_{\mathbb{S}^2}=n,\qquad \left<W\right>_{\mathbb{S}^2}=\frac{4\pi}{3},\qquad  \left<V\right>=-\frac{16\sqrt{2}}{15},
\eea
and where $\eta_1\eta_2=\pm1$ correspond to the sign in (\ref{bound}) and $\eta_1\eta_2 \,n=-|n|$.
The energy can be also calculated directly from the energy density $\mathcal{E}=2\mu^2\mathcal{U}$ 
\begin{eqnarray}
E=2\mu^2\int d^2xW\int dr r^2V=2\mu^2\left(\frac{4\pi}{3}\right)\left(\frac{16\sqrt{2}}{15}\frac{\lambda |n|}{\mu}\right),
\end{eqnarray}
what confirm the result (\ref{energia1}).
The potential $W(u\bar u)$ evaluated on the solution $u(\theta,\phi)$ is a function of  $\theta\in[0,\pi]$ given by
$
W(\theta)=\frac{1}{8}\left(3-\cos\theta\right)\sin^ 2\theta.
$
It takes its maximum value at $\theta=2\arctan(\sqrt{1+\sqrt{2}})$. The profile of $W(\theta)$ is sketched in Fig.\ref{don}.
Similarly, plugging $\xi(r)$ to $V(\xi)$ one gets a function  $V(r)=2\left(1-\frac{r^2}{R_0^2}\right)$ for $r\in [0,R_0]$. The energy density vanishes outside this interval.
The complex map $u$ covers the unit $\mathbb{S}^2$ sphere once and it has one zero for $\theta = 0$. Therefore the corresponding axially depending part of the energy density tends to 0 as we approach $\theta =0$ for any $r$. The resulting energy density vanishes on a semi-line $z \in [0,\infty)$ which makes the soliton similar to a half-doughnut.

%%%%%%%%%%%%%%%%%%%%%%%%%%%%%%%%%%%%%%%%%
\subsection{Baby skyrmions on a flat brane}\label{sec1}
%%%%%%%%%%%%%%%%%%%%%%%%%%%%%%%%%%%%%%%%%
A simple example can be found for the following potential
\begin{equation}
V(\xi)=1-\cos 2 \xi, \;\;\;\;  W(u\bar{u}) = \frac{|u|^2}{1+|u|^2} = \frac{1}{2}(1-n^3)
\end{equation}
where the potential $W(u\bar{u})$ takes the most typical form (usually referred as the old baby potential). It has minimum at $u=0$. The potential $V(\xi)$ is the simplest two vacua Skyrme potential (two vacua on the segment $[0,\pi]$ which must be covered as we search for the topological solitons with the baryon charge {\it i.e.}, skyrmions). \\
In cylindrical coordinates $x^1=r$, $x^2=\phi$, $x^3=z$ the BPS equations (\ref{bps1}) and (\ref{bps2}) became
\bea
&& \frac{2\lambda n}{r}\frac{ff_r}{(1+f^2)^{2}}=\eta_1\alpha \frac{f}{\sqrt{1+f^2}}\label{bps1a}\\
&&\sin^2\xi\,\xi_z=\frac{\eta_2}{\alpha}\sqrt{2}\mu\sin\xi\label{bps2a}
\eea
where $u=f(r)e^{in\phi}$ and $\xi=\xi(z)$. Equations (\ref{bps1a}) and (\ref{bps2a}) have constant  vacuum  solutions $(u,\xi)=\{(0,0),(0,\pi)\}$. 
The solution of (\ref{bps2a}) must cover the interval $[0,\infty)$, therefore we require $f(r)\rightarrow\infty$ for $r\rightarrow 0$. Moreover, the solution $f(r)$ must reach a zero value, what can be satisfied by the choice $\alpha=-\eta_1\,{\rm sign}(n)$. The solution of (\ref{bps1a}) is a compacton having a profile
\begin{equation}
f(r)=\left\{ 
\begin{array}{ll}
\frac{1-\frac{ r^2}{4\lambda |n|}}{\sqrt{1-\left(1-\frac{ r^2}{4\lambda |n|} \right)^2}} & r \leq 2\sqrt{\lambda |n|} \\
 & \\
0 &  r \geq 2\sqrt{\lambda |n|}
\end{array} \right.
\end{equation}
and the solution of (\ref{bps2a}) is a kink (anti-kink) for $\eta_1\eta_2\,{\rm sign}(n)=-1\,(+1)$ respectively. For instance, the anti-kink with topological charge $Q_{k}=-1$ is also a compact solution 
\begin{equation}
\xi(z) = \left\{ 
\begin{array}{ll}
\pi & z \leq z_0-\frac{1}{\sqrt{2}\mu} \\
\arccos [\sqrt{2} \mu (z-z_0)]  & z \in \left[ z_0-\frac{1}{\sqrt{2}\mu}, z_0+\frac{1}{\sqrt{2}\mu} \right] \\
0 &z  \geq z_0+\frac{1}{\sqrt{2}\mu}
\end{array} \right. 
\end{equation}
where $z_0$ is a free constant. The topological charge of the compact baby-skyrmion is given by the integral 
\be
Q_{bs}=\frac{i}{2\pi}\int_{K(0,R)}d^2x\frac{1}{r}\frac{\epsilon'_{ij}u_i\bar u_j}{(1+|u|^2)^2}=-n, 
\ee
where $K(0,R)$ is a disc with center at $r=0$ and the raduis $R=2\sqrt{\lambda|n|}$ whereas the kink/anti-kink topological charge takes the value $Q_{k}=-\eta_1\eta_2\,{\rm sign}(n)$, where $\eta_1\eta_2=\pm 1$ corresponds to signs in (\ref {bps0}) and $(\ref{bound1})$. It follows that the energy bound reads 
\be
E=2\lambda\mu |n| Q_{k} \left<\sqrt{W}\right>_K\left<\sqrt{V}\right>=\frac{32\pi}{9}\sqrt{2}\lambda \mu |n|,
\ee
where
\bea
&&\left <\sqrt{W}\right>_K=\frac{8}{3R^2}\int_{K(0,R)}d^2x\sqrt{W}=\frac{4\pi}{3},
\\
&&\left<\sqrt{V}\right>=\int_{-\frac{1}{\sqrt{2}\mu}}^{\frac{1}{\sqrt{2}\mu}} dz \sin^2(\xi)\xi_z\sqrt{V}=\frac{4\sqrt{2}}{3}Q_b.
\eea

The solution describes a compact baby-skyrmion located on a compact domain wall.
Another way to understand this solution is to look at it as a composition of a Skyrme string (a baby skyrmion with trivial $z$ dependence) and a Skyrme brane. Both objects, if treated separately, exist as vacuum solutions of the model (the vacuum manifold of the BPS Skyrme model is extremely large), so they correspond to zero energy excitations. However, their bound state has a finite mass.

%%%%%%%%%%%%%%%%%%%%%%%%%%%%%%%%%%%%%%%%%
\subsection{Baby-skyrmions with a discrete axial symmetry}\label{sec3}
%%%%%%%%%%%%%%%%%%%%%%%%%%%%%%%%%%%%%%%%%
The model has many interesting solutions when potential $W$ depends also on another two   components of the iso-vector $\vec n$ {\it i.e.} $n^1$ and $n^2$. In next two paragraphs we present some examples of solutions for such a potential. A characteristic feature of considered potential is that the axial continuous symmetry of the energy density factor $W(u,\bar u)$ is replaced by a discrete one. 

\subsubsection{Power function solutions}
 We shall discuss the following potentials
\begin{equation}
V(\xi)=1-\cos 2 \xi, \;\;\;\;  W(u,\bar u) = \left(\frac{1+n^ 3}{2}\right)^4\left|\frac{n^1+in^2}{1+n^3}+w_0^n\right|^{2(1-\frac{2\delta}{n})}
\end{equation}
where  $\delta>0$. We choose cylindrical coordinates and assume that fields depend on  coordinates in the following way $\xi=\xi(z)$, $u=u(r,\phi)=f(r)e^{in\phi}-w_0^n$, where $w_0=c\, e^{i\phi_0}$ is a constant complex number with $c$ and $\phi_0$ being real. The term $w_0\neq 0$ breaks down the axial continuous symmetry to a discrete one. The idea of such non-central potentials was given in  \cite{sawado}. Due to presence of a constant $w_0$ the potential $W$ is a function of $\phi$ too. Indeed,
\be
W(u,\bar u)=\frac{f(r)^{2(1-\frac{2\delta}{n})}}{(1+|u|^2)^4},
\ee
where $|u|^2=f(r)^2+c^{2n}-2c^n f(r)\cos{(n(\phi-\phi_0))}$.
The BPS equations take the form
\bea
&& \frac{2\lambda n}{r}\frac{ff_r}{(1+|u|^2)^{2}}=\eta_1\alpha \frac{f^{1-\frac{2\delta}{n}}}{(1+|u|^2)^2}\label{bps1c},\\
&&\sin^2\xi\xi_z=\frac{\eta_2}{\alpha}\sqrt{2}\mu\sin\xi\label{bps2c}.
\eea
where equation (\ref{bps1c}) involves in fact only a radial coordinate $r$.  For $\alpha=\eta_1$ the solution reads
\be
f(r)=\left(\frac{\delta}{2\lambda n^2}\right)^{\frac{n}{2\delta}}r^{\frac{n}{\delta}}.\label{powersol}
\ee
Note that for $\delta=1$ the field $u$ became a holomorphic function of a complex variable $re^{i\phi}$.
Unlike the solution in the section (\ref{sec1}) the baby-skyrmion found here is not  compact. However, the domain wall being a solution of (\ref{bps2c}) is still compact. The domain wall has a form of kink for $\eta_1\eta_2=1$ and anti-kink for $\eta_1\eta_2=-1$ with profile given by function $\xi(z)=\arccos [-\eta_1\eta_2\sqrt{2} \mu (z-z_0)]$ on $z \in  [z_0-\frac{1}{\sqrt{2}\mu}, z_0+\frac{1}{\sqrt{2}\mu}] $. 

%%%%%%%%%%%%%%
\begin{figure}[h!]
\centering
\subfigure[]{\includegraphics[width=0.32\textwidth]{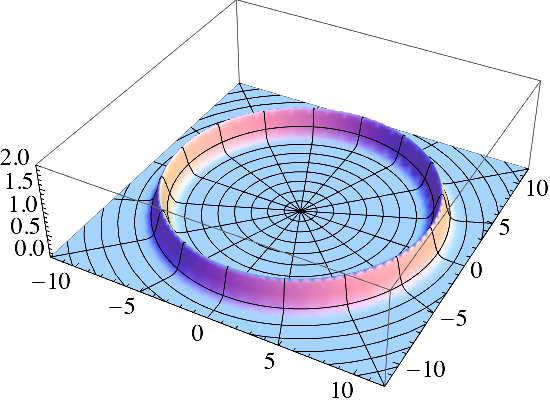}}\subfigure[]{\includegraphics[width=0.32\textwidth]{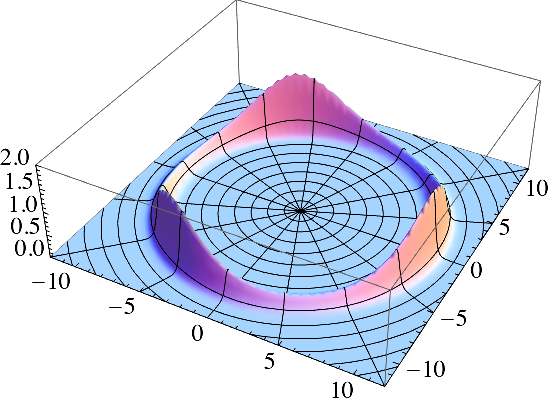}}\subfigure[]{\includegraphics[width=0.32
\textwidth]{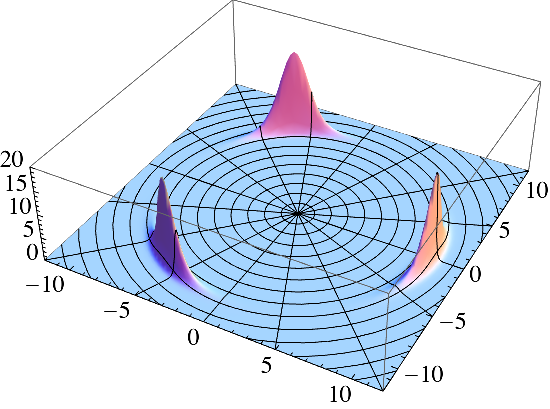}}
\caption{The energy density  $W(r,\phi)$ for the solution with $n=3$ and $\delta=0.2$ and for  $c=0.1$ (a),  $c=0.6$ (b), $c=1.0$ (c).}\label{ed1}
\end{figure}
%%%%%%%%%%%%%%
\begin{figure}[h!]
\centering
\subfigure[]{\includegraphics[width=0.32\textwidth]{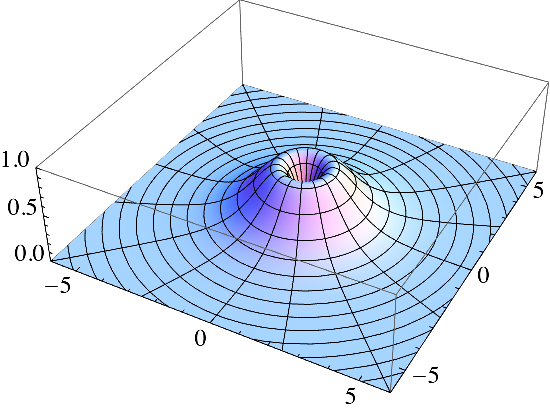}}\subfigure[]{\includegraphics[width=0.32\textwidth]{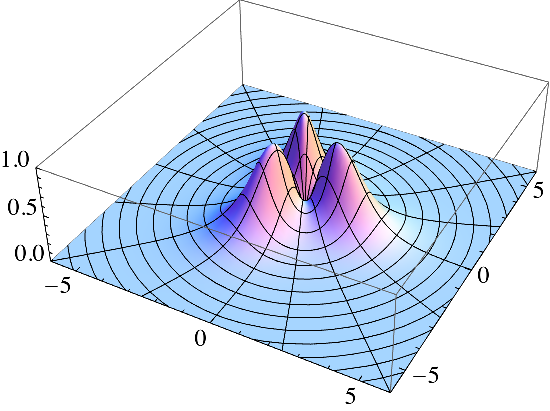}}\subfigure[]{\includegraphics[width=0.32
\textwidth]{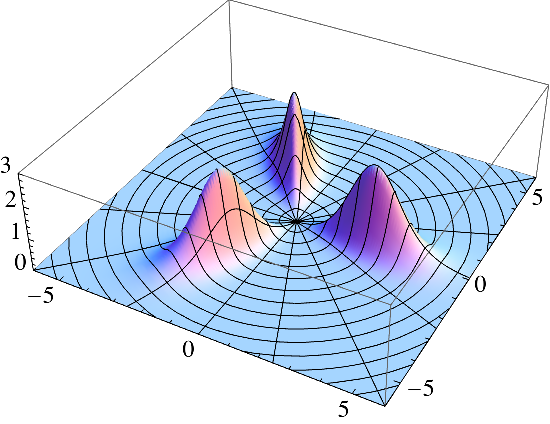}}
\caption{The energy density  $W(r,\phi)$ for the solution with $n=3$ and $\delta=3
.0$ and for  $c=0.1$ (a), $c=0.6$ (b), $c=1.0$ (c).}\label{ed2}
\end{figure}
%%%%%%%%%%%%%%

The bound of the energy is saturated by the BPS solutions and it reads
\bea
E=2\mu^2\int_{\mathbb{R}^2} d^2x W\int dz V=2\mu^2\left(\frac{4\sqrt{2}}{3\mu}\right)\int_{\mathbb{R}^2} d^2x W.
\eea
There are some restrictions on $\delta$ originated in requirement of finiteness of the integral $\int d^2x\,W$, namely, it must hold $0<\delta<2|n|$.  Some examples of the energy density for $n=3$ are sketched in Fig.\ref{ed1} and Fig.\ref{ed2}. The value of the parameter $c^n$ express a grade of breaking of the axial symmetry. For $c^n\rightarrow 0$ the energy density approaches the axially symmetric one. The energy density of solutions with $n>0$ in the limit $\delta \rightarrow 0$ coincide with the energy density of solutions with $n<0$ in the limit $\delta\rightarrow\infty$.
For axially symmetric configurations ($w_0=0$) the integral $\int d^2xW$ reads
\be
\int d^2xW=\lambda\frac{\pi^3}{3}\frac{\delta}{n^2}\frac{n^2-\delta^2}{\sin\left(\frac{\delta\pi}{|n|}\right)}
\ee 
and it is independent on ${\rm sign}(n)$. For $\delta\rightarrow |n|$ the integral simplifies to $\frac{2\pi}{3}\lambda|n|$. When $c^n$ is essentially different to zero the integral $\int d^2xW$ is not symmetric with respect to $n\rightarrow -n$. It follows that configurations which are not axially symmetric and differ only by the sign of $n$ have in general different energies.

\subsubsection{Generalization}
One can choose a potential $W$ in quite general form, namely 
\be
W(u,\bar u)=\left(\frac{1+n^ 3}{2}\right)^4|\psi|^2F^2(|\psi|),\label{Wgen}
\ee
where $F(f)$ is a non-negative valued function and $\psi$ is defined as follows
$$
\psi\equiv\frac{n^1+in^2}{1+n^3}+w_0^n=u+w_0^n.
$$
The potential (\ref{Wgen}) takes the form
\begin{eqnarray}
W=\frac{f^2(r)F^2(f)}{(1+|u|^2)^{4}},
\end{eqnarray}
where as before the numerator does not depend on variable $\phi$. It follows that the BPS equation reduces to two ordinary differential equations
\begin{eqnarray}
&&\frac{2\lambda n}{r}f_r=\eta_1\alpha F(f),\label{bps1d}\\
&&\sin^2\xi\xi_z=\frac{\eta_2}{\alpha}\mu\sqrt{1-\cos(2\xi)}\label{bps2d}
\end{eqnarray}
where we have skipped the common denominator $(1+|u|^2)^2$ on both sides of  (\ref{bps1d}).
The solution of equation (\ref{bps2d}) takes the form of a compact kink which interpolates between vacuum values $\xi=0$ and $\xi=\pi$ and has a profile $\xi(z)=\arccos{\left[-\frac{\eta_2}{\alpha}\sqrt{2}\mu(z-z_0)\right]}$. It is a kink for $\eta_2/\alpha>0$ and an anti-kink for $\eta_2/\alpha<0$. The auxiliary parameter $\alpha$ is fixed by requirement that $f(r)$ covers whole interval $[0,\infty)$. 

\begin{figure}[h!]
\centering
\subfigure[]{\includegraphics[width=0.32\textwidth]{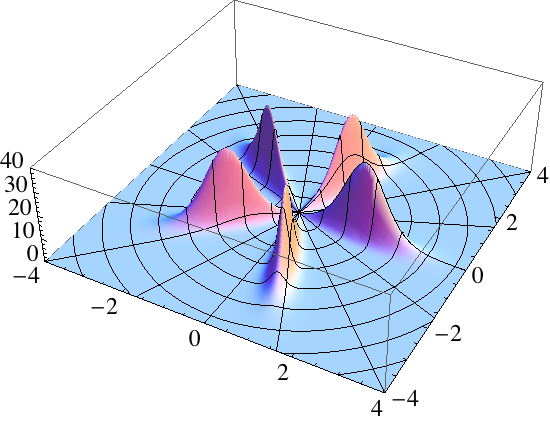}}\subfigure[]{\includegraphics[width=0.32\textwidth]{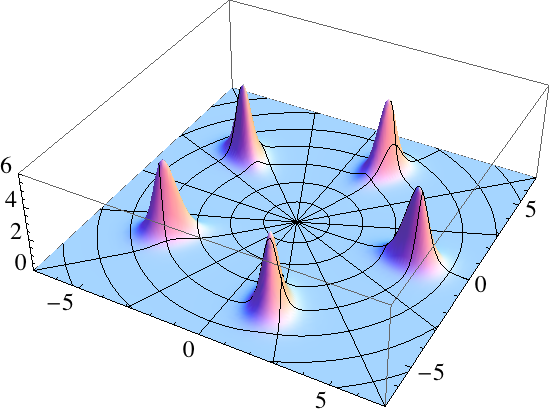}}\subfigure[]{\includegraphics[width=0.32
\textwidth]{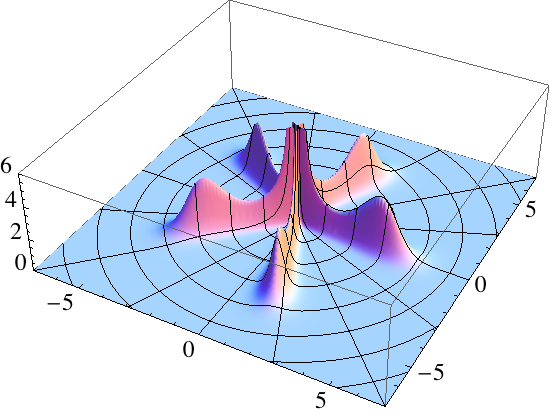}}
\caption{The energy density  $W(r,\phi)$ for the solution with $n=5$ and $c=1.03$ and for (a) $m=-\frac{1}{2}$, (b) $m=1$, (c) $m=\frac{1}{4}$.}\label{ed3}
\end{figure}

As example we shall consider $F(f)$ in the form
\be
F(f)=f^{1-\frac{1}{m}}\sqrt{1+f^{\frac{2}{m}}},
\ee
where $m$ is a free real parameter. Then the solution reads
\be
f(r)=\sinh^m\left(\frac{r^2}{4\lambda |m| |n|}\right),\label{solhyp}
\ee
where  $\alpha$ has been fixed as  $\alpha=\eta_1\,{\rm sign}(m)\,{\rm sign}(n)$. Such solution is obviously not a power like function. Many other solutions are possible for different choice of $F(f)$. Examples of the energy density for solution (\ref{solhyp}) are shown in Fig.\ref{ed3}.
%%%%%%%%%%%%%
%%%%%%%%%%%%%

%%%%%%%%%%%%%%%%%%%%%%%%%%%%%%%%%%%%%%%%%
\subsection{Skyrmion as a three-brane defect}\label{sec4}
%%%%%%%%%%%%%%%%%%%%%%%%%%%%%%%%%%%%%%%%%

A solution being an intersection of three-domain walls can be obtained in the parametrization of $U$ given by
\begin{eqnarray}
U=\mathbb{I}\,\xi^4+i\xi^a\tau_a\qquad\qquad a=1,2,3
\end{eqnarray}
where one has to impose the condition $(\xi^1)^2+(\xi^2)^2+(\xi^3)^2+(\xi^4)^2=1$ which implies $U^{\dagger} U=\mathbb{I}$. There are only three independent scalar fields. One can parametrise a 3-sphere in a usual way
\begin{eqnarray}
\xi^1&=&\sin\theta_1\sin\theta_2\sin\theta_3\nonumber\\
\xi^2&=&\sin\theta_1\sin\theta_2\cos\theta_3\nonumber\\
\xi^3&=&\sin\theta_1\cos\theta_2\nonumber\\
\xi^4&=&\cos\theta_1
\end{eqnarray}
where  $\theta_1,\theta_2\in[0,\pi]$ and $\theta_3\in[0,2\pi]$.
The independent fields $\phi^1$, $\phi^2$, $\phi^3$  can be introduced by a stereographic projection of a 3-sphere on a 3-dim Euclidean  hyperplane 
\begin{eqnarray}
\vec\phi\equiv\left(   \begin{array}{c} % brackets may be (...), [...], \{...\}, or left out
      \phi^1  \\
      \phi^2  \\
      \phi^3  \\
   \end{array}\right)=\frac{1}{1+\xi^4}\left(   \begin{array}{c} % brackets may be (...), [...], \{...\}, or left out
      \xi^1  \\
      \xi^2  \\
      \xi^3  \\
   \end{array}\right).
\end{eqnarray}
In terms of $\phi^a$ the fields  $\xi^a, \xi^4$ are given by expressions 
\begin{eqnarray}
\xi^a=\frac{2\phi^a}{1+\vec\phi^2},\qquad\qquad\xi^4=\frac{1-\vec\phi^2}{1+\vec\phi^2},\qquad\qquad a=1,2,3,
\end{eqnarray}
The expression $\mathcal{B}^{\mu}$ introduced in (\ref{top-curr}) is given in terms of $L_{\mu}\equiv U^{\dagger}\partial_{\mu}U$ 
\begin{eqnarray}
L_{\mu}&=&i\epsilon'_{ijk}(\xi^i\partial_{\mu}\xi^j)\tau_k+i\left[\xi^4\partial_{\mu}\xi^k-\xi^k\partial_{\mu}\xi^4\right]\tau_k\nonumber\\
&=&\frac{2i}{(1+\vec\phi^2)^2}\left[2(\vec\phi\times\vec\phi_{\mu})\cdot\vec\tau+(1-\vec\phi^2)(\vec\phi_{\mu}\cdot\vec\tau)+2(\vec\phi\cdot\vec\phi_{\mu})(\vec\phi\cdot\vec\tau)\right],
\end{eqnarray}
where $\vec\phi_{\mu}\equiv\partial_{\mu}\vec\phi$ and it reads
\begin{eqnarray}
\mathcal{B}^{\mu}=\frac{2}{3\pi^2}\frac{\epsilon'^{\mu\nu\alpha\beta}}{(1+\vec\phi^2)^6}\left[(1-\vec\phi^2)^3(\vec\phi_{\nu}\times\vec\phi_{\alpha})\cdot\vec\phi_{\beta}+6\left(3+(\vec\phi^2)^2\right)(\vec\phi\cdot\vec\phi_{\beta})(\vec\phi_{\nu}\times\vec\phi_{\alpha})\cdot\vec\phi \right].
\end{eqnarray}
In this section the coordinates $x^{\mu}$ are Cartesian. We shall consider the ansatz $\phi^a=\phi^a(x^a)$ {\it i.e.} each field depends on exactly one  spatial coordinate. It follows that there is only one non-vanishing component of $\mathcal{B}^{\mu}$, namely 
\begin{eqnarray}
\mathcal{B}^{0}=\frac{4}{\pi^2}\frac{\phi'^1\phi'^2\phi'^3}{(1+\vec\phi^2)^3},
\end{eqnarray}
where $\phi'^a(x^a)\equiv\frac{d\phi^a(x^a)}{dx^a}$. The BPS equations (\ref{Bogomolny}) take the following form
\begin{eqnarray}
4\lambda\frac{\phi'^1\phi'^2\phi'^3}{(1+\vec\phi^2)^3}\pm\mu\sqrt{\mathcal{U}}=0.
\end{eqnarray}
The expression $\vec\phi^2$ depends on all three spatial coordinates $x^1,x^2,x^3$. For this reason one can choose the potential in such a way that a common denominator cancels out on both sides of the BPS equation. A quite general potential with this property can be chosen in the form
\begin{eqnarray}
\mathcal{U}=\left[\frac{F_1(\phi^1)F_2(\phi^2)F_3(\phi^3)}{(1+\vec\phi^2)^3}\right]^2,
\end{eqnarray}
where functions $F_i(\phi^i)\ge 0$ shall be specified below. In such a case the BPS equations can be decomposed in three equations, each for one field $\phi^a$
\begin{eqnarray}
\phi'^1&=&\eta_1\alpha F_1(\phi^1),\label{eqbps1}\\
\phi'^2&=&\eta_2\beta F_2(\phi^2),\label{eqbps2}\\
\phi'^3&=&\frac{\eta_3}{\alpha\beta}\frac{\mu}{4\lambda}  F_3(\phi^3),\label{eqbps3}
\end{eqnarray}
 where $\alpha,\beta>0$ are some auxiliary constants and signs $\eta_k$ are such that $\eta_1\eta_2\eta_3=\mp 1$. In particular, a symmetric choice has the form $
 \alpha=\beta=\sqrt[3]{\frac{\mu}{4\lambda}}$.
Note that functions $F_a(\phi^a)$ can be chosen in many topologicaly non-equivalent ways. 
 
\begin{figure}[h!]
\centering
\subfigure[]{\includegraphics[width=0.5\textwidth]{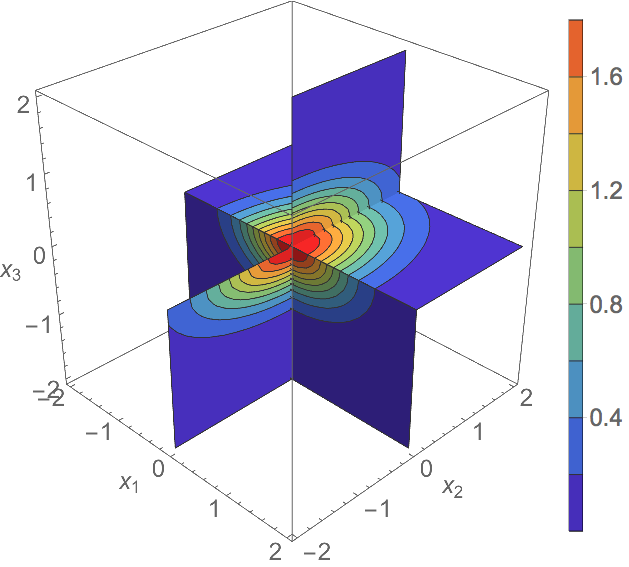}}\subfigure[]{\includegraphics[width=0.5\textwidth]{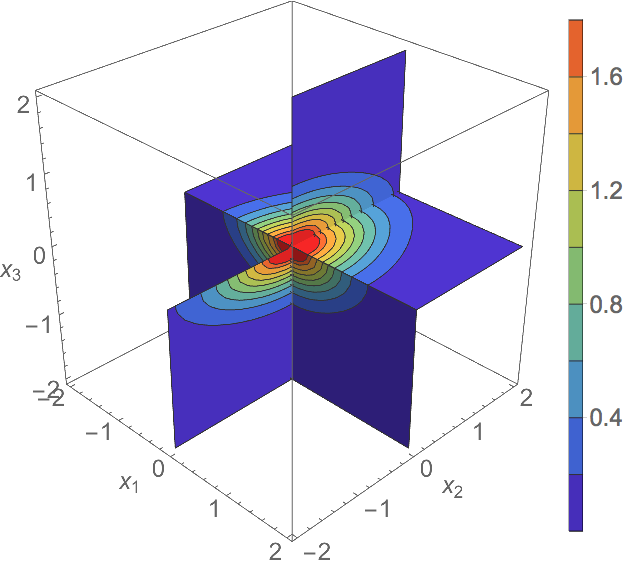}}\caption{The energy density as the function of $x^1$, $x^2$ and $x^3$ for a BPS configuration. Figure (a) correspond to the case where potential  has no zeros; total energy reads $E=\frac{7}{8}\pi^2\approx8.636$. Figure (b) correspond to potential leading to kinks; total energy reads $E=5.968$.}\label{kinks}
\end{figure}

 The simplest choice is $F_a=const$. In such a case the potential has no zeros and vacuum manifold is like that for the Skyrme potential. Then, the solution must cover whole 3-sphere. It is possible if $\phi^a$ interpolate between $-\infty$ and $+\infty$. For instance, taking $F_a(\phi^a)=1$ one gets solution
 \begin{eqnarray}
 \phi^1=\eta_1\alpha (x^1-x^1_0),\qquad  \phi^2=\eta_2\beta (x^2-x^2_0),\qquad\phi^3=\frac{\eta_3}{\alpha\beta}\frac{\mu}{4\lambda} (x^3-x^3_0),
 \end{eqnarray}
 where $x^a_0$ are some arbitrary constants. A picture of the energy density for such a solution is shown in Fig.\ref{kinks}a. The baryon charge defined by $B=\int d^3x\mathcal{B}^0$ takes value $B=-1$ for a considered solution.

 On the other side, when the potential has zeros the vacuum manifold is non-trivial. In particular for appropriate form of potential there exist kink/anti-kink solutions. An example of such solutions is given for the potential defined by functions
 \begin{eqnarray}
F_a(\phi^a):=\cos^2(\phi^a)\qquad a=1,2,3.
\end{eqnarray}
Then equations (\ref{eqbps1}), (\ref{eqbps2}) and (\ref{eqbps3}) are solved by 
 \begin{eqnarray}
 \phi^1(x^1)&=&\arctan(\eta_1\alpha (x^1-x^1_0)),\\
 \phi^2(x^2)&=&\arctan(\eta_2\beta (x^2-x^2_0)),\\
 \phi^3(x^3)&=&\arctan\left(\frac{\eta^3}{\alpha\beta}\frac{\mu}{4\lambda} (x^3-x^3_0)\right),
 \end{eqnarray}
 which have a form of kinks interpolating between vacua $\pm\frac{\pi}{2}$. Topological charges of kinks are given by some integer numbers $Q_a=\eta_a$, however, they have nothing to do with the baryon number $B$ which in fact is not anymore a good number because it is not even integer for a considered potential.  The energy density of pertinent solution is shown in Fig.\ref{kinks}b. In both examples solutions are plotted for the following choice of parameters
$$
 \eta_1=1,\quad\eta_2=1,\quad\eta_3=-1,\quad\lambda=2,\quad\mu=1,\quad\alpha=\frac{1}{2},\quad\beta=\frac{1}{3}.
$$ 
 
%%%%%%%%%%%%%%%%%%%%%%%%%%%%%%%%%%%%%%%%%
\section{Pressure and Equation of State}
%%%%%%%%%%%%%%%%%%%%%%%%%%%%%%%%%%%%%%%%%
For any potential $\mathcal{U}$ the BPS Skyrme model is a field theory describing a perfect fluid. This means that the energy-momentum tensor has the characteristic form, which for static configurations reads
\bea
T^{00}&=&\varepsilon = \lambda^2\pi^4 \mathcal{B}_0^2+\mu^2 \mathcal{U}, \\
T^{ij}&=&p\,\delta^{ij}= \left( \lambda^2\pi^4 \mathcal{B}_0^2-\mu^2 \mathcal{U} \right)\delta^{ij}
\eea
where $\varepsilon$ and $p$ denote, respectively, energy density and pressure. From the conservation law of the energy-momentum tensor we find that the pressure must have a constant value.  Furthermore, as shown in \cite{term} the constant pressure equation is a first integral of the full second order static field equations. Hence, non-zero pressure solutions for any symmetry breaking potential are given by the following equation
\be
\lambda^2\pi^4 \mathcal{B}_0^2-\mu^2 \mathcal{U} = p.
\ee
%%%%%%%%%%%%%%%%%%%%%%%%%%%%%%%%%%%%%%%%%%%%%
\subsection{The isospin symmetry breaking as a continuous deformation of the potential}

The models investigated in previous sections are given from the very beginning as the models with the isospin symmetry broken. It is particularly interesting  to have the model where the isospin symmetry  breaking can be introduced as a continuous deformation of the model that possesses the isospin symmetry. 
For this reason we shall consider the BPS Skyrme model with the potential which can be smoothly deformed. In analogy to section \ref{sec2} we assume dependence of fields on spherical coordinates, namely $\xi=\xi(r)$ and $u=v(\theta)e^{in\phi}$. The BPS equation (\ref{bps0}) can be cast in the form 
\begin{eqnarray}
\left[\frac{2}{\sin\theta}\frac{v_{\theta}v}{(1+v^2)^2}\right]\left[\lambda n\sin^2\xi\,\frac{\xi_r}{r^2}\right]=\pm\mu\sqrt{W(v)}\sqrt{\mathbb{V}(\xi)},\label{bps02}
\end{eqnarray}
where we shall use a letter $\mathbb{V}$ for the potential and keeping a letter $V$ for volume. We shall choose the potential $W(|u|)$ as the function which depends on the parameter $\varepsilon\in[0,1]$
\begin{eqnarray}
W(|u|)= (1+\varepsilon)^2-2\varepsilon(1+n^3)=(1+\varepsilon)^2-\frac{4\varepsilon}{1+|u|^2}.\label{potW}
\end{eqnarray}
For $\varepsilon=0$ the potential $W=1$, what leads to the model with the presence of the isospin symmetry. The other extremal value $\varepsilon=1$ leads to the model with the potential $W=2(1-n^3)=4\frac{|u|^2}{1+|u|^2}$ which has been studied before in the context of baby-skyrmion on a spherical brane. One has to stress the fact that the potential (\ref{potW}) taken for $\varepsilon=1$ is qualitatively different from the other cases  $\varepsilon\in[0,1)$. The reason is that the value of the potential at the minimum $|u|=0$ is $W(|u|=0)=0$ for $\varepsilon=1$ whereas $W(|u|=0)>0$ for $\varepsilon\in[0,1)$ so the vacuum manifold is qualitatively different in both cases.

We are interested in solutions $v(\theta)$ that satisfy the boundary conditions $v(0)=0$ and $v(\pi)=\infty$. In order to simplify equation (\ref{bps02}) we define new variable $x:=\cos\theta$ and new field component $w(x):=(1+v^2)^{-1}$. The $r$-dependent component can be simplified with help of variable $z:=\frac{2\mu r^3}{3\lambda|n|}$ and new field $\zeta(z):=\frac{1}{2}(\xi-\frac{1}{2}\sin{2\xi})$.  The equation (\ref{bps02}) simplifies to the form
\begin{eqnarray}
2w_x\zeta_z=\pm{\rm sign}(n)\left[\left(1+\varepsilon\right)^2-4\varepsilon w\right]^{1/2}\sqrt{\mathbb{V}(\zeta)}.\label{bps03}
\end{eqnarray}
Let us observe that the equation $2w_x=\sqrt{W}$ has solution
\begin{eqnarray}
w(x)=\frac{1+x}{2}\left[1+\varepsilon\frac{1-x}{2}\right],\label{solw}
\end{eqnarray}
which gives
\begin{eqnarray}
v(\theta)=\sqrt{\frac{1}{w(\cos\theta)}-1}\qquad {\rm where}\qquad w(\cos\theta)=\cos^2\frac{\theta}{2}\left[1+\varepsilon\sin^2\frac{\theta}{2}\right].\label{solv}
\end{eqnarray}
The solution (\ref{solv}) became $v(\theta)=\tan\frac{\theta}{2}$ for $\varepsilon=0$ and $v(\theta)=\frac{\tan\frac{\theta}{2}\sin\frac{\theta}{2}}{\sqrt{1+\sin^2\frac{\theta}{2}}}$ for $\varepsilon=1$ in concordance with our previous considerations, see (\ref{u0}) and (\ref{v1}). Then the BPS equation (\ref{bps03}) takes the form
\begin{eqnarray}
\zeta_z=\pm{\rm sign}(n)\sqrt{\mathbb{V}(\zeta)}.
\end{eqnarray}
Many particular properties of the solution $\zeta(z)$ depend on the choice of the potential $\mathbb{V}(\zeta)$. Since we are interested in studying models in presence of pressure then we choose the potential $\mathbb{V}(\zeta)=\zeta^{\beta}$ and compare our results with those presented in the paper \cite{term} in absence of the isospin symmetry breaking.

%%%%%%%%%%%%%%%%%%%%%%%%%%%%%%%%%%%%%%%%%%%%
\subsection{Non-zero pressure case}
\begin{figure}[h!]
\centering
\subfigure[]{\includegraphics[width=0.3\textwidth]{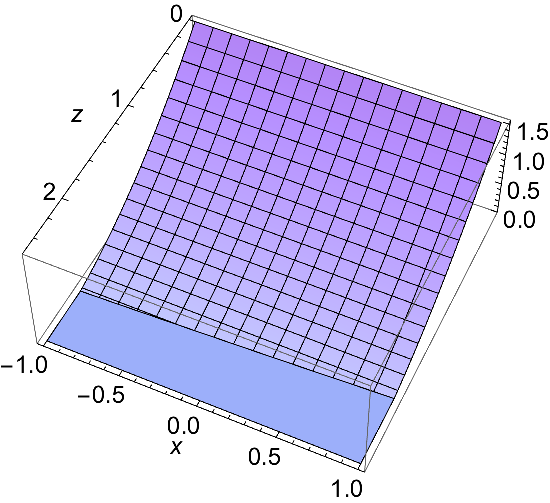}}\subfigure[]{\includegraphics[width=0.3\textwidth]{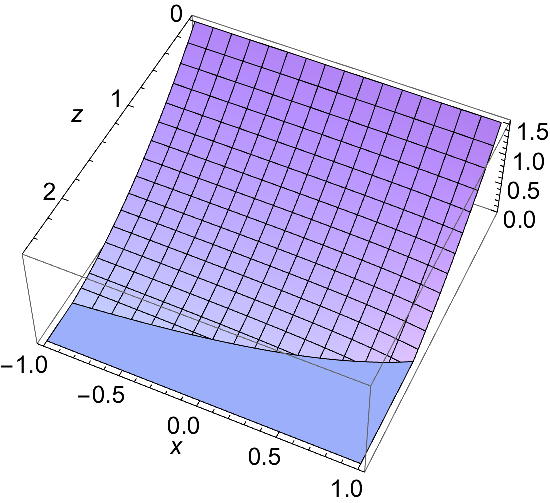}}\subfigure[]{\includegraphics[width=0.3\textwidth]{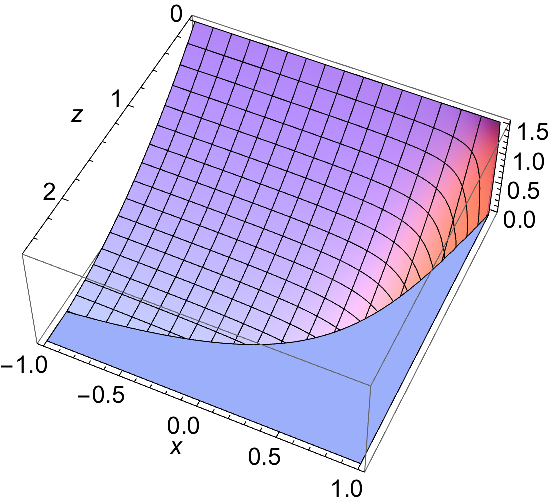}}
\caption{The function $\zeta(z,x)$ for $\tilde p=0.1$ and for (a) $\varepsilon=0.1$, (b) $\varepsilon=0.5$, (c) $\varepsilon=1.0$.}\label{zeta}
\end{figure}

When pressure is taken in account then the BPS equation must be substituted by the following one
\begin{eqnarray}
\pi^2\lambda\mathcal{B}^0\pm\mu\sqrt{\mathcal{U}+\tilde p}=0,\label{eqpress}
\end{eqnarray}
where $\tilde p\equiv p/\mu^2$. One could not expect that for generic situation there works  any obvious separation ansatz which allows to reduce equation (\ref{eqpress}) to the product of two equations. However, one can to check what happens for the simplest choice {\it i.e.} when fields do depend on spherical coordinates in the following way $\xi=\xi(r,\theta)$ and $u=v(\theta)e^{in\phi}$. Since $\vec n$, or equivalently $u$, do not depend on coordinate $r$ then $\mathcal{B}^0$ has the same functional form as for the case with $\xi=\xi(r)$. The only difference is that $\xi(r)$ is replaced by $\xi(r,\theta)$. In terms of new coordinates $x$ and $z$ equation (\ref{eqpress}) became
\begin{eqnarray}
2\partial_xw(x)\partial_z\zeta(z,x)=\pm{\rm sign}(n)\sqrt{W(w)\mathbb{V}(\zeta)+\tilde p}.\label{eqpressure}
\end{eqnarray}
We shall chose the potential $W(w)$ in the form (\ref{potW}) and $\mathbb{V}(\zeta)=\zeta^{\beta}$ in analogy to \cite{term}. Plugging (\ref{solw}) to (\ref{eqpressure}) one gets equation for $\zeta(z,x)$. It has a form
\begin{eqnarray}
\partial_z\zeta(z,x)=\pm{\rm sign}(n)\sqrt{\zeta^{\beta}+\frac{\tilde p}{(1-\varepsilon x)^2}}\label{eqzeta}
\end{eqnarray}
for $\varepsilon\in[0,1)$. For $\varepsilon=1$ the potential $W(w(x))=(1-x)^2$ takes value zero at $x=1$. Consequently $\partial_xw=0$ at $x=1$ so according to (\ref{eqpressure}) one can expect singularity in derivative of $\partial_z\zeta$ for $\tilde p\neq 0$. Indeed, such behaviour of the solution is present, see Fig.\ref{zeta}c. We shall impose the following boundary conditions on field $\xi$
\begin{eqnarray}
\xi(r=0,\theta)=\pi\qquad \xi(r=\infty,\theta)=0,
\end{eqnarray}
or equivalently on field $\zeta$
\begin{eqnarray}
\zeta(z=0,x)=\frac{\pi}{2}\qquad \zeta(z=\infty,x)=0.
\end{eqnarray}
Such conditions can be imposed on solution of (\ref{eqzeta}), howewer, one has to choose as well signs in a way that $\pm{\rm sign}(n)= -1$. We shall also fix the power $\beta=1$ since systematic study of all possible potentials $\zeta^{\beta}$ is out of scope of this paper.  It follows that equation (\ref{eqzeta}) has exact solution
\begin{eqnarray}
\zeta(z,x)=\frac{1}{4}\left[2\pi+z^2-2z\sqrt{2\pi+\frac{4\tilde p}{(1-\varepsilon x)^2}}\right].
\end{eqnarray}
In fact it is quite interesting result because in absence of an obvious separation ansatz one would expect only numerical solutions of resulting PDE for the model with broken the isospin symmetry. Here we were able to give an exat solution. Let us comment about one important point related to this solution. It is important to notice that the choice (\ref{solw}) is not unique. Since the model has a large symmetry (volume preserving diffeomorphisms) many alternative solutions can be mapped one to each other using such transformations. For this reason we do not have to worry about existence of multiple choice and just study the simplest one.

%%%%%%%%%%%%%%%%%%%%
%%%%%%%%%%%%%%%%%%%%

It follows that $\zeta$ does not depend on $x$ for $\tilde p=0$ and/or $\varepsilon=0$. In similarity to \cite{term} the solution is compact and the border of this compacton is determined by solution of the equation $\zeta(Z,x)=0$ giving
\begin{eqnarray}
Z(x)=\frac{2}{1-\varepsilon x}\left[\sqrt{\tilde p+\frac{\pi}{2}(1-\varepsilon x)^2}-\sqrt{\tilde p}.\right].
\end{eqnarray}
The main difference is that it clearly depends on $x$ for $\tilde p\neq0$. The radial coordinate that describes the compacton border behaves as $R\sim\sqrt[3]{Z}$.
The function $\zeta(z,x)$ is shown in  Fig.\ref{zeta}. For $\tilde p=0$ the  expression $Z$ became  $Z=\sqrt{2\pi}$. For the model with $\varepsilon=1$ the function $Z(x)\rightarrow0$ as $x\rightarrow 1$ what can be seen from Fig.\ref{zeta}c. and expansion at $x=1$
\begin{eqnarray}
Z(x)=\frac{\pi}{2\sqrt{\tilde p}}(1-x)+\mathcal{O}\left((1-x)^3\right).
\end{eqnarray}

\begin{figure}[h!]
\centering
\subfigure[]{\includegraphics[width=0.4\textwidth]{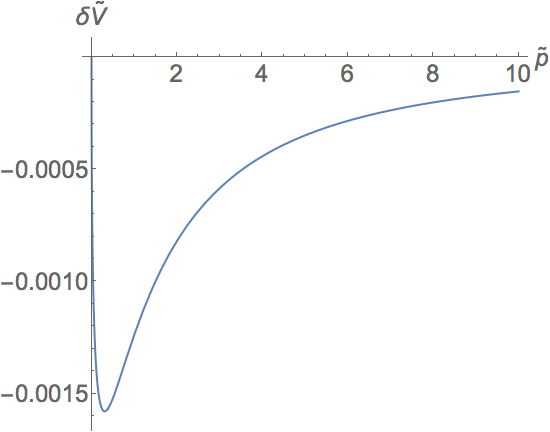}}\hskip 1cm\subfigure[]{\includegraphics[width=0.4\textwidth]{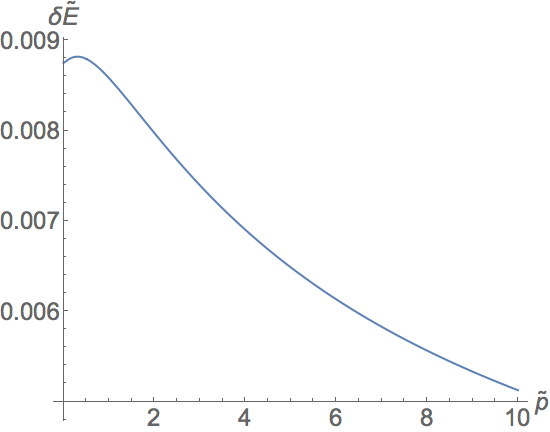}}\caption{(a) The volume difference $\delta\tilde V=\tilde V_{\varepsilon=0.1}-\tilde V_{\varepsilon=0}$  and (b) the energy difference $\delta\tilde E=\tilde E_{\varepsilon=0.1}-\tilde E_{\varepsilon=0}$  in dependence on pressure $\tilde p$.}\label{difference}
\end{figure}

\begin{figure}[h!]
\centering
\subfigure[]{\includegraphics[width=0.4\textwidth]{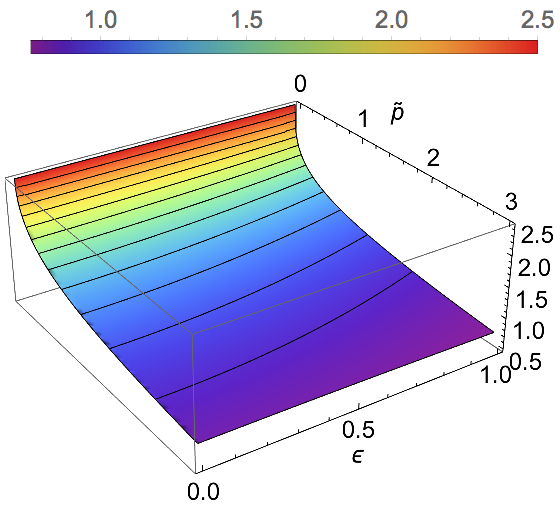}}\subfigure[]{\includegraphics[width=0.4\textwidth]{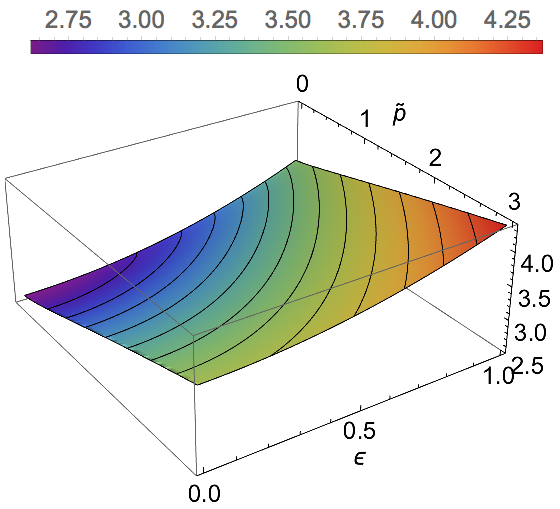}}\caption{ The volume (a) $\tilde V(\tilde p,\varepsilon)$ and the energy (b) $\tilde E(\tilde p,\varepsilon)$ of the compacton.}\label{state1}
\end{figure}

 \noindent
 A volume occupied by the compacton is given by the integral over base space
\begin{eqnarray}
V=\int_{\Omega} \mbox{vol}_{\mathbb{R}^3} =\int_{0}^{2\pi}d\phi\int_{-1}^{1}dx\int_{0}^{R(x)}r^2dr=\frac{\pi\lambda|n|}{\mu}\int_{-1}^1dx \int_{0}^{Z(x)}dz=\frac{2\pi\lambda|n|}{\mu}\tilde V.\label{Vbase}
\end{eqnarray}
where  $\tilde V\equiv \frac{1}{2}\int_{-1}^1dxZ(x)$, or with help of relations $\frac{dw}{\sqrt{W}}=\frac{1}{2}dx$ and $\frac{\sqrt{W}d\zeta}{\sqrt{\mathcal{U}+\tilde p}}=-dz$ as the integral on target space. In such a case one gets
\begin{eqnarray}
\tilde V=\int_{0}^1dw\int_0^{\pi/2}\frac{d\zeta}{\sqrt{\mathcal{U}+\tilde p}}\label{intV}.
\end{eqnarray}
An explicit expression for $\tilde V$ has the form
\begin{eqnarray}
\tilde V=\frac{\sqrt{\tilde p+\frac{\pi}{2}(1+\varepsilon)^2}-\sqrt{\tilde p+\frac{\pi}{2}(1-\varepsilon)^2}}{\varepsilon}-\frac{\sqrt{\tilde p}}{\varepsilon}\ln\left[\frac{\sqrt{\tilde p}+\sqrt{\tilde p+\frac{\pi}{2}(1+\varepsilon)^2}}{\sqrt{\tilde p}+\sqrt{\tilde p+\frac{\pi}{2}(1-\varepsilon)^2}}\right].\label{eqstate}
\end{eqnarray}
One can deduce from (\ref{eqstate}) that  the dominant change of a compacton volume induced by weak $\varepsilon\ll 1$ effects of the isospin symmetry breaking is proportional to  the second power of the parameter $\varepsilon$   
\begin{eqnarray}
\tilde V=2\left[\sqrt{\tilde p+\frac{\pi}{2}}-\sqrt{\tilde p}\right]+\left[\frac{ \tilde p}{\sqrt{\tilde p+\frac{\pi}{2}}}-\frac{2}{3}\left(\sqrt{\tilde p}+\frac{\tilde p^2}{2(\tilde p+\frac{\pi}{2})^{3/2}}\right)\right]\varepsilon^2+\mathcal{O}(\varepsilon^3).
\end{eqnarray}
The second term of expansion is a negative valued function which behaves as $-\frac{2}{3}\tilde p^{1/2}$ for small pressure and as $-\frac{\pi^2}{16}\tilde p^{-3/2}$ for high pressure. The function has its minimum value $\approx-0.157$ for $\tilde p\approx0.309$. It follows that for slightly broken the isospin symmetry the volume change effects in order $\varepsilon^2$ are significant only if pressure is not too high or extremely small. In Fig.\ref{difference}a we present  difference $\delta\tilde V=\tilde V_{\varepsilon}-\tilde V_{\varepsilon=0}$ which represent the total effect of the volume change due to isospin symmetry breaking. For  chosen value of $\varepsilon$ there is virtually no difference between total value $\delta\tilde V$ and the second term of expansion.
In Fig.\ref{state1}a  we present  the rescaled volume of the compacton $\tilde V$ in dependence on pressure $\tilde p$ and deformation parameter $\varepsilon$. One can conclude from this picture that in order to maintain a constant value of compacton volume one has to decrease pressure as increasing $\varepsilon$.

Let us shortly discuss the energy and the energy density. The energy of the compacton contains the potential part and the pressure part
\begin{eqnarray}
E=\mu^2\int d^3x(2\mathcal{U}+\tilde p)=2\pi\lambda\mu|n|\tilde E\qquad{\rm where}\qquad\tilde E=\frac{1}{2}\int_{-1}^1dx\int_{0}^{Z(x)}dz(2\mathcal{U}+\tilde p).\label{en1}
\end{eqnarray}
%%%%%%%%%%%%%%%%%%%%
\begin{figure}[h!]
\centering
\subfigure[]{\includegraphics[width=0.4\textwidth]{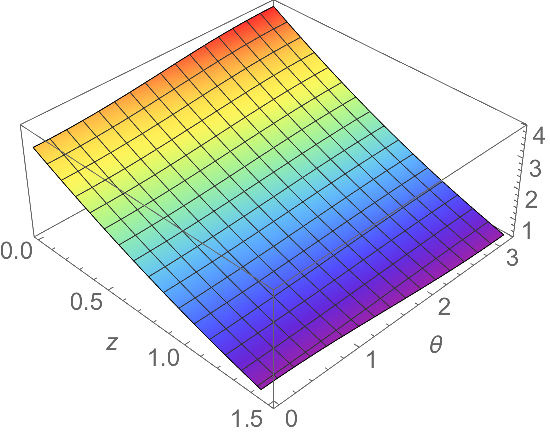}}
\subfigure[]{\includegraphics[width=0.4\textwidth]{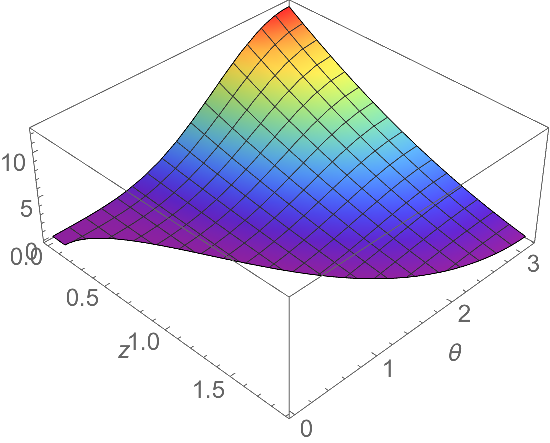}}
\subfigure[]{\includegraphics[width=0.4\textwidth]{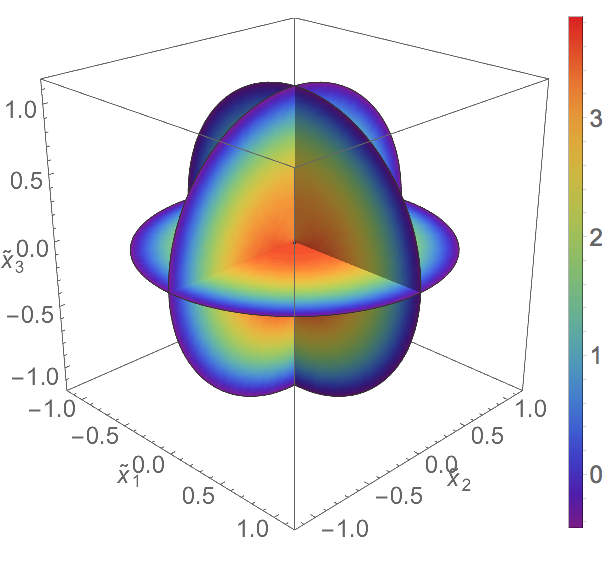}}
\subfigure[]{\includegraphics[width=0.4\textwidth]{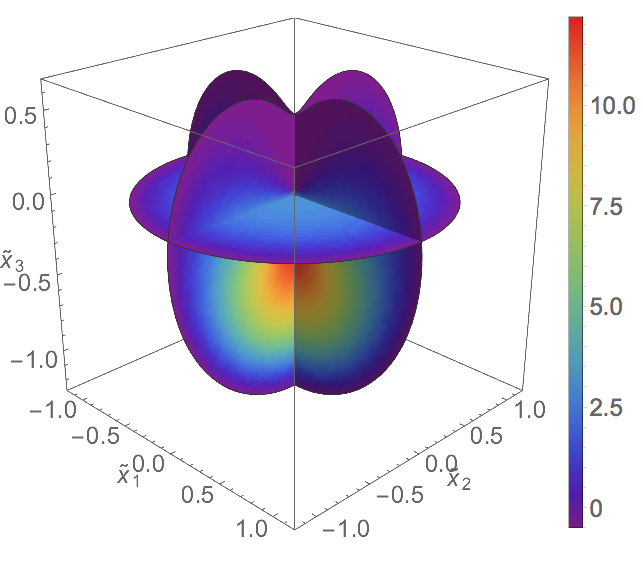}}
\caption{The expression $2\mathcal{U}+\tilde p$ for $\tilde p=0.5$ and (a), (c) $\varepsilon=0.05$, and (b), (d) $\varepsilon=0.95$.}\label{dens}
\end{figure}
%%%%%%%%%%%%%%%%%%%%
\noindent 
In Fig.\ref{dens} we present the expression $2\mathcal{U}+\tilde p$ which is proportional to the energy density. The rescaled coordinates $\tilde x^k$ are are given by $\tilde x^k=\sqrt[3]{\frac{2\mu}{3\lambda|n|}}x^k$.  Clearly the dependence of the energy density on $\theta$ is an effect of the isospin symmetry breaking. The grade of asymmetry in the diagram grows with increasing of $\varepsilon$. The compacton border is defined as $\zeta(Z,\cos(\theta))=0$ and therefore the energy density is constant there. The value of this constant is just $\tilde p$ because $\mathcal{U}=0$ at the border. The energy (\ref{en1})  can be  also expressed as the integral on a target space
\begin{eqnarray}
\tilde E=\int_{0}^1dw\int_{0}^{\pi/2}d\zeta\frac{2\mathcal{U}+\tilde p}{\sqrt{\mathcal{U}+\tilde p}}\label{intE}
\end{eqnarray}
where the integral (\ref{intE}) reads
\begin{eqnarray}
\tilde E&=&\frac{1}{9\varepsilon}\left[\left[\pi(1+\varepsilon)^2-\tilde p\right]\sqrt{\tilde p+\frac{\pi}{2}(1+\varepsilon)^2}-\left[\pi(1-\varepsilon)^2-\tilde p\right]\sqrt{\tilde p+\frac{\pi}{2}(1-\varepsilon)^2}\right.\nonumber\\
&+&\left.3\tilde p^{3/2}\ln\left(\frac{\sqrt{\tilde p}+\sqrt{\tilde p+\frac{\pi}{2}(1+\varepsilon)^2}}{\sqrt{\tilde p}+\sqrt{\tilde p+\frac{\pi}{2}(1-\varepsilon)^2}}\right)\right].\label{tildeE}
\end{eqnarray}
This expression has following expansion for small values of $\varepsilon$
\begin{eqnarray}
\tilde E=\frac{2}{3}\left[\tilde p^{3/2}+(\pi-\tilde p)\sqrt{\tilde p+\frac{\pi}{2}}\right]+\frac{1}{9}\left[2p^{3/2}+\frac{\pi^3+3\pi\tilde p(\pi-\tilde p)-4\tilde p^3}{2\left(\tilde p+\frac{\pi}{2}\right)^{3/2}}\right]\varepsilon^2+\mathcal{O}(\varepsilon^3).\label{exp2}
\end{eqnarray}

%%%%%%%%%%%%%%%%%%
%%%%%%%%%%%%%%%%%%

%%%%%%%%%%%%%%%%%%
%%%%%%%%%%%%%%%%%%

The dominant contribution to the energy for slightly broken the isospin symmetry is proportional to square of $\varepsilon$. The pertinent coefficient is a function which behaves as $\frac{\sqrt{3}}{9}\pi^{3/2}+\frac{2}{9}\tilde p^{3/2}$ for $\tilde p$ close zero and as $\frac{3\pi^2}{16}\tilde p^{-1/2}$ for high pressure. It follows that the variation of the total energy caused by effect of the isospin symmetry breaking is non negligible if pressure is not too high. The coefficient proportional to $\varepsilon^2$ takes its maximum value $\approx0.882$  for $\tilde p\approx0.309$. The plot of difference $\delta \tilde E=\tilde E_{\varepsilon}-\tilde E_{\varepsilon=0}$ is sketched in Fig.\ref{difference}b. It became a very good approximation of the second term of (\ref{exp2}) for  $\varepsilon=0.1$. The variation of the energy of compacton caused by the isospin symmetry breaking is less significant for higher pressures. 
In Fig.\ref{state1}b we present diagram of rescaled energy $\tilde E$ as a function of  pressure $\tilde p$ and parameter $\varepsilon$. It follows from analysis of curves $\tilde E=const$ that increasing of $\varepsilon$ leads to decreasing of pressure. 

\begin{figure}[h!]
\centering
\includegraphics[width=0.5\textwidth, height=0.4\textwidth]{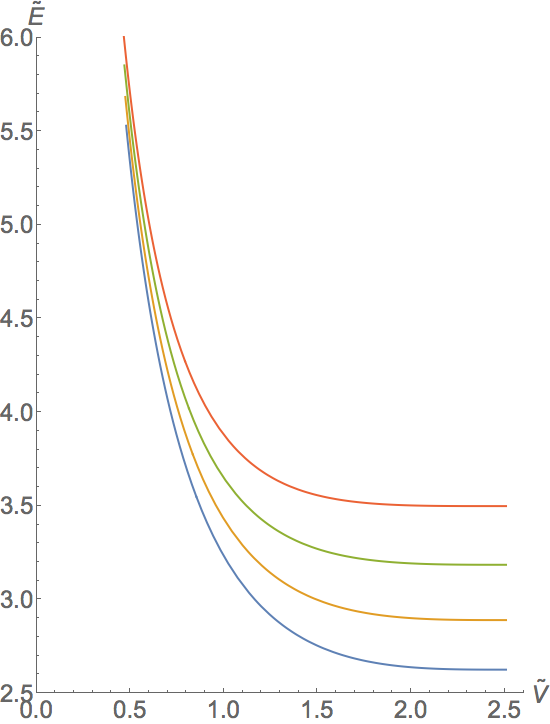}\caption{The energy of the compacton as a funcion of its volume $\tilde E(\tilde V)$. From down to up the curves correspond to $\varepsilon=0.01,\,\varepsilon=0.55,\,\varepsilon=0.8,\,\varepsilon=0.999$.}\label{state2}
\end{figure}

In Fig.\ref{state2} we plot the function $\tilde E(\tilde V)$. The upper value of $\tilde V$ is limited by the value $\sqrt{2\pi}\approx 2.506$ ($\tilde p=0$) and the lower value corresponding to pressure $\tilde p=10$ changes monotonically between values $\approx 0.478$ for $\varepsilon=0.01$ and $\approx 0.464$ for $\varepsilon=0.999$. Compactons in the model with broken the isospin symmetry have in general higher energy comparing with energy of compactons occupying the same volume but being solutions of the model possessing the isospin symmetry. The value of the energy calculated for fixed $\tilde V$ increases with increasing of $\varepsilon$.

%%%%%%%%%%%%%%%%%%
%%%%%%%%%%%%%%%%%%

\section{Conclusions}
In this paper we have studied the BPS Skyrme model with a potential which explicitly breaks the isospin symmetry. We have shown that the strong isospin symmetry breaking gives origin to new type of exact solutions that have a form of composite skyrmions. Such solutions are obtained from the BPS equation which admits separation in a set of ordinary first order equations each one for each field. We gave examples of three different types of such composite structures. In sec. \ref{sec2} we constructed the compact baby-skyrmion on a spherical compact brane. The bound state of such two objects has finite mass although each one treated separately exist as a vacuum solution of the BPS model. Other possibility have been explored in sec. \ref{sec1}.  This composite skyrmion is a composition of a baby-skyrmion located on a compact domain wall. We have shown that the class of such solutions is very large and in fact involves also solutions with azimuthal angle dependence, sec. \ref{sec3}. It leads to very interesting situation when the energy density is not axially symmetric but rather possesses only a residual discrete axial symmetry. 
An alternative possibility of having a composite skyrmion has been explored in \ref{sec4} where we have constructed such object as  intersection of three branes. The topological charge of such a solution is $B=\pm1$. We have studied two qualitatively different cases: potential has or does not have additional zero.  In the later case, the topological charge is given by a baryon number of field configuration and we found usual BPS skyrmions. On the other hand, when the potential has new zeros, the vacuum manifold is nontrivially deformed, which can support some other topological solutions. In our example each individual scalar field has a kink(anti-kink) solution. The topological charge of such configuration is a product of topological charges of kinks. 
\\
All these examples show some similarities with previously known effects, where for example skyrmions have been also constructed as a composite state of intersecting domain walls \cite{nitta1}.

Another problem associated with the isospin symmetry breaking is an understanding how this phenomenon affects thermodynamic properties of  composite skyrmions. In particular we have studied corrections to the (energy-pressure and volume-pressure) equation of state in the weak and strong symmetry breaking limit. In a generic case, the symmetry breaking becomes less significant for higher pressure. This is an expected result as in the high pressure limit the role of the potential becomes in fact immaterial, while it is the derivative part (sextic term) which governs the thermodynamical properties \cite{eos}. On the other hand, the symmetry breaking modifies equation of state close to the equilibrium ($P=0$). This may be of some relevance for astrophysical applications of the BPS Skyrme model. For example, neutron stars mass-radius relation will be very weekly affected by such a modification of the potential - the week symmetry breaking contribution is further suppressed by high pressure.

There are several directions in which the present work can be continued. First of all, one should go beyond the classical regime and include the semiclassical corrections together with the Coulomb interaction.   

Secondly, one can ask how classical properties of skyrmions depend on the isospin breaking if the usual (perturbative) pionic part of the model $\mathcal{L}_{sk}$ is included. Furthermore, such a modification of the potential should have an impact on dynamical properties of skyrmions as their rotation \cite{hab} and collision \cite{dave}. Let us remark that some time-dependent properties of the BPS skyrmions have been recently studied in \cite{arpad}. 

There is also a way to include the medium effects from the isospin symmetry breaking. This can be easily achieved by promoting the usual time derivative to a proper covariant derivative \cite{medium}. However, this leads to a non-BPS theory which probably complicates computations for higher baryon numbers.

Since physical nuclei do not possess perfect spherical or axial symmetry we expect that BPS skyrmions without such symmetries can be more adequate for a realistic description of nuclear matter.  In this sense, the isospin symmetry breaking seems to qualitatively improve the applicability of the BPS Skyrme model. Undoubtedly, this should be verified once the semiclassical effects have been taken into account.

\section*{Acknowledgments}
The author is indebted to Andrzej Wereszczynski for discussion and many valuable comments.

\end{document}